\documentclass[conference]{IEEEtran}
\IEEEoverridecommandlockouts
\usepackage{amsmath,amsfonts}
\usepackage{algorithmic}
\usepackage{graphicx}
\usepackage{multirow}
\usepackage{textcomp}
\usepackage{cases}
\usepackage{tabularx}
\usepackage{xcolor}
\usepackage{booktabs}
\usepackage{subfig}
\usepackage{tikz}
\usepackage[hidelinks]{hyperref}

\newcommand\copyrighttext{%
  \footnotesize \textcopyright 2021 IEEE. Personal use of this material is permitted.
  Permission from IEEE must be obtained for all other uses, in any current or future
  media, including reprinting/republishing this material for advertising or promotional
  purposes, creating new collective works, for resale or redistribution to servers or
  lists, or reuse of any copyrighted component of this work in other works.
  DOI: \href{https://doi.org/10.1109/CloudIntelligence52565.2021.00011}{https://doi.org/10.1109/CloudIntelligence52565.2021.00011}}
\newcommand\copyrightnotice{%
\begin{tikzpicture}[remember picture,overlay]
\node[anchor=south,yshift=10pt] at (current page.south) {\fbox{\parbox{\dimexpr\textwidth-\fboxsep-\fboxrule\relax}{\copyrighttext}}};
\end{tikzpicture}%
}

\DeclareMathAlphabet{\altmathcal}{OMS}{cmsy}{m}{n}

\begin{document}

\title{Learning Dependencies in Distributed Cloud Applications to Identify and Localize Anomalies}

\author{
\IEEEauthorblockN{Dominik Scheinert,
Alexander Acker,
Lauritz Thamsen,
Morgan K. Geldenhuys, and
Odej Kao}
\IEEEauthorblockA{Technische Universit{\"a}t Berlin, Germany, \{firstname.lastname\}@tu-berlin.de}
}

\maketitle
\copyrightnotice

\begin{abstract}
Operation and maintenance of large distributed cloud applications can quickly become unmanageably complex, putting human operators under immense stress when problems occur.
Utilizing machine learning for identification and localization of anomalies in such systems supports human experts and enables fast mitigation.
However, due to the various inter-dependencies of system components, anomalies do not only affect their origin but propagate through the distributed system.
Taking this into account, we present \emph{Arvalus} and its variant \emph{D-Arvalus}, a neural graph transformation method that models system components as nodes and their dependencies and placement as edges to improve the identification and localization of anomalies.
Given a series of metric KPIs, our method predicts the most likely system state - either normal or an anomaly class - and performs localization when an anomaly is detected.
During our experiments, we simulate a distributed cloud application deployment and synthetically inject anomalies. 
The evaluation shows the generally good prediction performance of Arvalus and reveals the advantage of D-Arvalus which incorporates information about system component dependencies.
\end{abstract}


\section{Introduction}
\label{sec:introduction}
The evolution of IT systems enables rapid innovation in a variety of fields like medicine, autonomous transportation or manufacturing.
Often, applications and services are deployed in cloud computing environments, referred to as distributed cloud applications or microservices~\cite{wang2018cloudranger}.
Furthermore, they often are decomposed into lightweight, self-contained and independently deployable services which enables high degrees of isolation, fine-grained scaling and fast adaption of individual components to customer requirements.
However, these deployments are reaching sizes of hundreds to thousands of interconnected system components~\cite{dragoni2017microservices}.
Such complex distributed systems are hard to operate and human experts are increasingly in need of support and automation.

To guarantee high availability and low latency, inevitably occurring problems must be timely identified and localized.
Significant effort was done to research the detection of system anomalies by utilizing machine learning (ML) methods~\cite{Gulenko2016,mariani2018localizing,Nedelkoski2019}.
Thereby, monitoring is employed to collect key performance indicator (KPI) metrics such as network latency or resource utilization from all relevant components.
Based on the collected data, models are trained to detect anomalies.
Although these methods provide support for the operation of monolithic applications, their use in largely distributed and interconnected environments leads to problems.
Occurring anomalies in a single component can propagate through the system.
In such cases, employing anomaly detection leads to many alarms from the different affected components, leaving the search for the origin to human operators.

Fault localization - sometimes referred to as root cause analysis (RCA) - in distributed systems has been addressed in several works~\cite{wang2018cloudranger, sauvanaud2016anomaly, li2019generic}. 
However, most of these methods are relying on manual rules, selectively constructed metrics, or assume access to source code, input data or configuration parameters.
Another line of work explicitly operates on the dependencies of services, modelling them as correlation or causality graphs~\cite{Wu2020microRCA}. 
Here, the localization of anomalies is done by backtracking graph edges, resulting in a ranked list of possible locations.
Yet, the mining of the graphs from existing data and the backtracking when anomalies occur is difficult.
During graph mining, the calculation of KPI inter-dependencies usually has a quadratic computational complexity, posing a limit to the scale of systems.
Moreover, backtracking with change-based heuristics on calculated KPI inter-dependencies is not generally applicable to all possible system anomalies.

To overcome these limitations, we propose our data-driven anomaly identification and localization approach \emph{Arvalus} and its variant \emph{D-Arvalus} which utilizes a novel graph convolution (GC) method to model system component inter-dependencies.
Meta-information about the system deployment are used to build an initial graph structure.
A trainable node feature extraction and edge feature transformer are employed to complement the graph.
D-Arvalus applies GC to aggregate node features of neighboring nodes.
Finally, anomaly identification and localization is done based on the calculated node features.
Instead of heuristics and rules, this method introduces a trainable and therefore general anomaly identification and localization.
Our approach is evaluated in a simulated environment.
The deployment structure of this environment is based on the deployment layout of real cloud applications.
To evaluate our proposed graph convolution method, Arvalus and its variant D-Arvalus are compared.


The remainder of the paper is structured as follows:
\autoref{sec:relatedWork} discusses the related work with regards to RCA in distributed systems.
\autoref{sec:approach} describes our approach and explains the idea of exploiting the dependencies of distributed systems for improved anomaly identification and localization.
\autoref{sec:evaluation} presents our experimental setup and our evaluation results. 
\autoref{sec:conclusion} concludes the paper.
\section{Related Work}
\label{sec:relatedWork}
Anomaly detection based on statistics and machine learning is a widely studied field.
Ahmed et. al~\cite{Ahmed2016ASO} provide a taxonomy of anomaly detection methods based on classification, statistics, information theory, and clustering. 
Due to the propagation of anomalies in distributed and interdependent systems, the application of anomaly detection methods alone eventually leads to a large number of false alarms.
Therefore, the related field of root cause analysis (also referred to as fault localization) receives much interest.
In~\cite{zhou2019latent} the authors propose a method for fault localization using system traces via supervised model training.
Unlike Arvalus, their method requires manually selected features. 
Likewise, the authors of~\cite{meng2020localizing} motivate for the use of a path condition time series algorithm in combination with an adapted random walk algorithm.
This not only allows to capture sequential relationships in time series data, but also to incorporate causal relationships, temporal order, and priority information of monitoring data.
The drawback of this method is its complexity.
With Arvalus we simplify the feature extraction by only considering temporal relationships within extracted slices of the respective time series. 
Additionally, we learn the characteristics of anomalies whereas their approach uses thresholds on root cause metrics which usually requires expert knowledge to be set correctly. 
In~\cite{sauvanaud2016anomaly}, the authors employ a Random Forest algorithm for anomaly detection and RCA in Virtual Network Functions. 
Similar to Arvalus, they carry out an analysis individually for each system component as well as an ensemble analysis.
The latter only considers the global prediction probabilities and neglects the underlying graph structure of system dependencies.   

Arvalus is within a class of approaches specific to distributed cloud systems which models system state in the form of a directed graph to store information and identify dependencies between the various nodes. 
A certain number of these are concerned with diagnosis after an anomaly has been reported and not the detection of anomalies themselves as is the case with our approach. 
\emph{MonitorRank}~\cite{10.1145/2494232.2465753} is a real time metric collection system used to perform RCA in service-oriented architectures. 
It proposes an unsupervised and heuristic (clustering) approach for producing a ranked list of possible root causes which aid monitoring teams in investigations. 

Further causality graph-based approaches aim to provide a system for end-to-end anomaly detection and root cause analysis. 
An approach called \emph{LOUD}~\cite{mariani2018localizing} trains a model with correct executions only and uses graph centrality algorithms to localize faulty resources in cloud systems. 
Although the authors claim that exclusively relying on positive training is able to detect and localize exceptions to the steady state, this does however prevent the identification of specific anomalies and therefore reduces the ability to respond to them appropriately.
\emph{GRANO}~\cite{wang2019grano} is an enterprise level software framework developed at Ebay consisting of a detection layer, an anomaly graph layer, and an application layer which assists fault resolution teams. 
This extensible framework employs multiple techniques for detecting and localizing anomalous behaviors for cloud-native distributed data platforms. 
In contrast, the authors of~\cite{Wu2020microRCA} propose to infer root causes by correlating application performance symptoms with corresponding system resource utilization.
Here, anomaly detection is realized using \emph{BIRCH}~\cite{gulenko2018detecting}, and RCA is implemented based on an attributed graph that models anomaly propagation across systems.



\section{Approach}
\label{sec:approach}
Our method utilizes KPIs of individual cloud system components like hosts, virtual machines or microservices to identify and localize anomalies.
Therefore, we model distributed cloud applications as graphs where each node represents a system component and edges represent dependencies between them.
The approach consists of three steps.
First, a KPI subseries of a component is transformed into a feature vector for the respective graph node.
Second, the feature vectors of neighboring nodes together with meta-information are used to calculate edge weights to realize the graph convolution operation.
Third, the feature vector of each node is utilized to identify and localize anomalies.
Besides a formal problem definition, the subsequent sections provide detailed explanations of these steps.



\subsection{Preliminaries}
\label{sec:approach_prel}

KPIs such as tracing or metric data are the basis for detecting and localizing anomalies in distributed cloud applications. 
When continuously collected over time, they provide an abstract representation of the state of each system component.

Metric data KPIs can be defined as multivariate time series, i.e. a temporally ordered sequence of vectors $S = ({S}_t \in \mathbb{R}^d : t=1,2,\ldots, T)$, where $d$ is the number of KPIs and $T$ defines the last sample time stamp.
For $S^{a}_{b}=(S_a, S_{a+1}, \ldots, S_b)$, we denote indices $a$ and $b$ with $a \leq b$ and $0 \leq a,b \leq T$ as time series boundaries in order to slice a given series $S^{0}_{T}$ and acquire a subseries $S^{a}_{b}$.

Distributed cloud applications can be modelled as graphs to represent the dependencies between system components.
A directed, weighted and attributed graph $G=(V,E)$ with $n$ nodes consists of a set of vertices $V=\{v_1, \ldots, v_n\}$ and a set of edges $E\subseteq \{(v_i,v_j)| v_i,v_j \in V\}$. Each node $v_i$ has a node feature vector $\Vec{x}_i \in \mathbb{R}^{F}$.
An edge $e_{ij} \Leftrightarrow (v_i,v_j)\in E$ describes a directed connection between vertex $v_i$ and $v_j$. 
Thus, the node $v_j$ is then called a neighbor of node $v_i$, formally written as $j\in \altmathcal{N}(i)$. 
Each edge $e_{ij}$ is attributed by a vector $\Vec{r}_{ij}$, containing meta information about the connection.
The adjacency matrix $A$ of a graph $G$ is an $n \times n$ matrix with entries $A_{ij}$ such that $A_{ij}=\Vec{r}_{ij}$ if an edge $e_{ij}$ exists, otherwise 0.

\subsection{Node Feature Extraction}
\label{sec:approach_extraction_representation}
\begin{figure}[t!]
    \centering
    \includegraphics[page=1,keepaspectratio,width=\columnwidth]{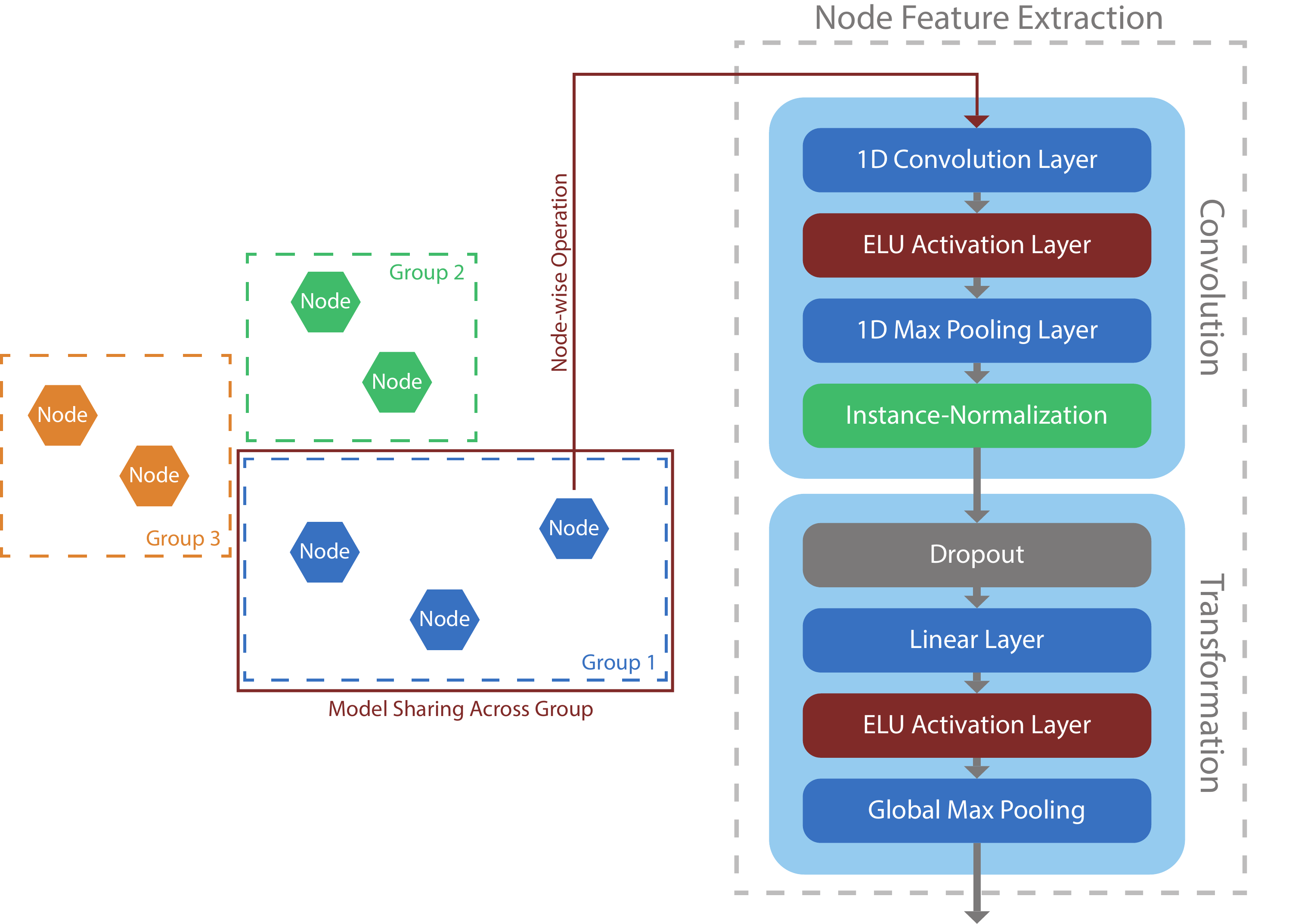}
    \caption{Node feature extraction consisting of a convolution and transformation block. It transforms KPI subseries of each node into a node feature vector. Hexagons symbolize the extracted KPI subseries.}
    \label{fig:arvalus_step1}
\end{figure}

The previously presented preliminaries enable us to describe our approach.
First, all replicas of the same application service are grouped.
We refer to them as service groups.
Other than that, no grouping is applied.
Next, our method transforms KPI subseries $S^{a}_{b} \in \mathbb{R}^{d\times(b-a)}$ of each node into a node feature vector $\Vec{x} \in \mathbb{R}^{F}$.
Intuitively, the resulting feature vector can be interpreted as a compact representation of the node state.
The steps are further visualized in \autoref{fig:arvalus_step1}.  

Initially, the \emph{Convolution} block aims at the extraction of features from the KPI subseries.
A convolution layer with $k$ filters, each of different filter size to capture features at different scale, is applied and the results are concatenated resulting in a feature matrix $X\in \mathbb{R}^{d \times ((b-a) \cdot k)}$.
After that we perform an ELU activation, 1D max-pooling and instance normalization.
The feature matrix dimensionality is preserved.
These steps aim at the extraction of meaningful features from the raw KPI subseries.
The \emph{Transformation} block transforms the feature matrix into a node feature vector.
During training, a dropout layer is used to mitigate overfitting.
A linear transformation is employed to map the feature matrix to a desired dimensionality $X'\in \mathbb{R}^{d\times F}$.
Finally, a column-wise global max-pooling results in a compact node representation $\Vec{x}\in \mathbb{R}^{F}$.

These operations are applied for each node.
However, every node group (circles of same colour enclosed by the dotted rectangle in~\autoref{fig:arvalus_step1}) shares the same model layer weights.

\subsection{Dependency Model}
\label{sec:approach_graph}

\begin{figure}[b!]
    \centering
    \includegraphics[page=2,keepaspectratio,width=\columnwidth]{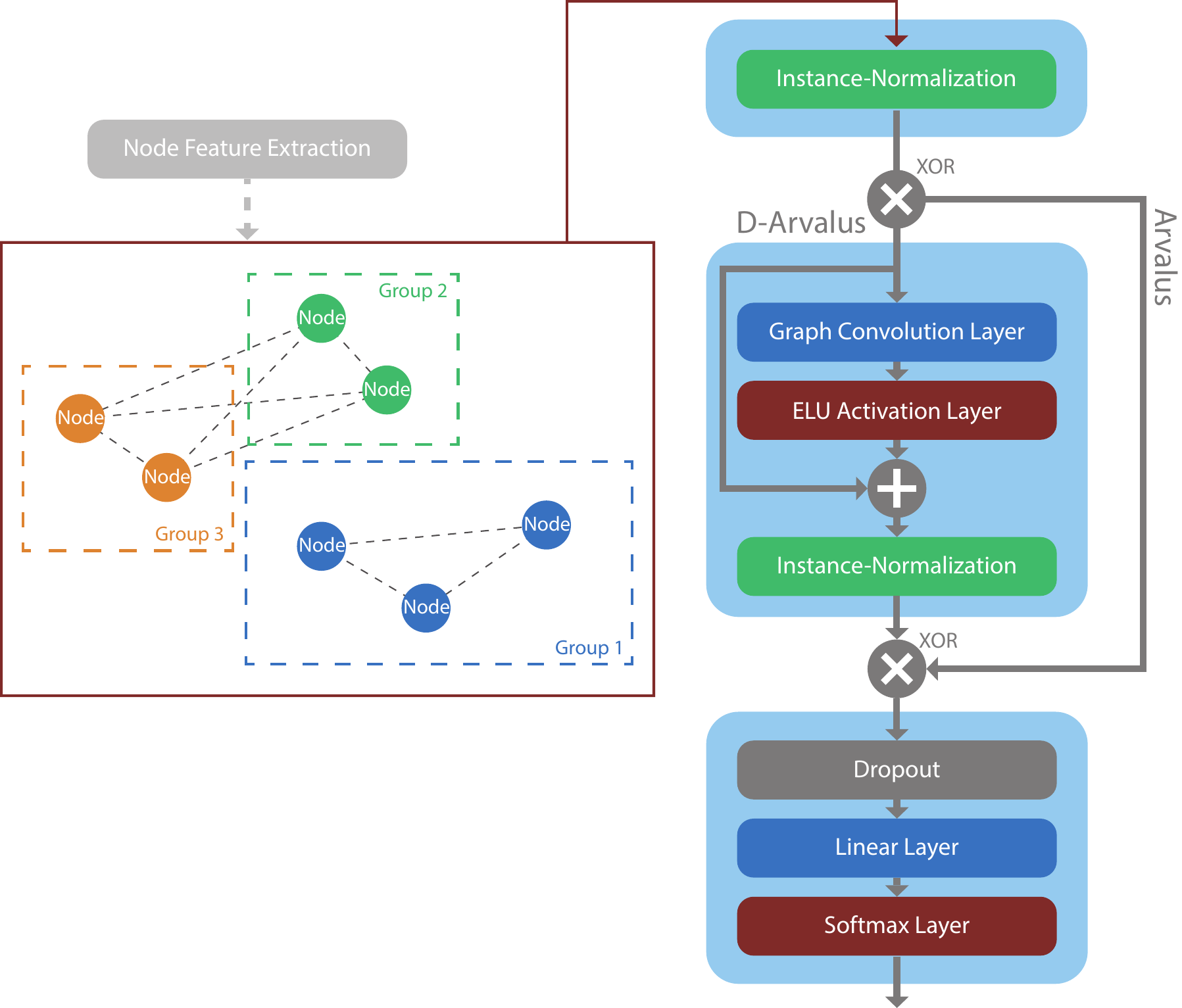}
    \caption{Dependency model: The graph with its learnable edge weights is used during GC to aggregate features of neighboring nodes. Circles symbolize the computed node feature vectors.}
    \label{fig:arvalus_step2}
\end{figure}

Distributed cloud system components have diverse inter-dependencies.
Virtual machines are hosted on hypervisor nodes, containers which are hosted on virtual machines contain microservices.
Thus, the state of a single system component is determined both by local executions and external factors. 
This is our motivation to model the underlying dependencies of such systems via edge weight learning and graph convolution.

Graph convolutional neural networks (GCNNs) are a class of neural networks which incorporates concepts from graph theory and aim at generalizing the convolution operation to be applied in non Euclidean domains~\cite{Zhang2019}.
The convolution operation therefor operates on the neighborhood of each graph node. 
GCNN methods can be roughly clustered into spectral and spatial methods. 
In this work, we focus on spacial methods since they are essentially defining the graph convolution in the vertex domain by leveraging the graph structure and aggregating node information from the neighborhoods in a convolutional fashion. 

\begin{table}
\centering
\caption{List of defined edge types.}
\small
    \begin{tabular}[t]{c||l}
        \toprule
        \textbf{Edge Type} & \textbf{Condition}\\
        \midrule
        \verb|Identity| & self-loop, for each node\\
        \midrule
        \verb|Host-Guest| & system $i$ is hosting system $j$\\
        \midrule
        \verb|Guest-Host| & system $i$ is hosted by system $j$\\
        \midrule
        \verb|Group| & system $i$ and system $j$ are part of same group\\
        \midrule
        \verb|Dependency| & any other uncovered kind of dependency\\
        \bottomrule
    \end{tabular}
\label{tbl:approach_edge_types}
\end{table}

We assign the previously calculated node features to the respective nodes and apply an instance normalization, respectively for each graph.
After that, Arvalus provides two operation modes, respectively referred to as Arvalus and D-Arvalus and conceptually depicted in \autoref{fig:arvalus_step2}.
It can either use the node features directly to predict and localize system anomalies in the last block (Arvalus) or it utilizes graph convolution to aggregate feature vectors of neighboring nodes, adding this information via a residual connection to the original node features (D-Arvalus).

Inspired by~\cite{velivckovic2018graph}, we propose our graph convolution approach designed for distributed cloud applications.
First, we construct an initial adjacency matrix $A$ of our graph by fabricating edges $e_{ij}$ between two nodes $v_i$ and $v_j$ using the edge types derived in \autoref{tbl:approach_edge_types}.
As defined in \autoref{sec:approach_prel}, directed edges are used.
The first three edge types are self-explanatory.
The fourth edge type introduces a unique edge tag for each existing component group (e.g. one for $Group1$, one for $Group2$, etc.). 
To reflect inter-dependencies that are not described by the the first four edge types (e.g. back-end service group depends on the identity service group to authorize access to certain database entries), the fifth edge type is used.
Whenever dependencies between two service groups exist,  we use a tag for each direction, e.g. $Group1\rightarrow Group2$ and $Group2\rightarrow Group1$. 
We argue that edges between graph nodes can be inferred from information about the respective deployment itself as presented in~\cite{wu:hal-02968710}.
Next, each edge is tagged with their respective edge type, as described above.
We apply a one-hot-encoding of the tags and attach them to the respective edges as edge attribute vectors. 
For $z$ unique tags, this results in edge attribute vectors $\Vec{r}_{ij} \in \mathbb{N}_0^z$.


Before applying the graph convolution, we transform each edge attribute vector into a scalar weight value.
More precisely, we formulate a trainable transformation function $f$ that uses the node features of two neighbor nodes $\Vec{x}_{i}$ and $\Vec{x}_{j}$ combined with the one-hot encoded edge vector $\Vec{r}_{ij}$.
Applying an ELU activation and posterior normalization results in an adjusted adjacency matrix $\Tilde{A}$ with
\begin{equation}
\Tilde{A}_{ij} = \frac{\exp\big(\text{ELU}\big(f(\Vec{x}_{i}, \Vec{r}_{ij}, \Vec{x}_{j})\big)\big)}{\sum\limits_{k\in \altmathcal{N}(i)}\exp\big(\text{ELU}\big(f(\Vec{x}_{i}, \Vec{r}_{ik}, \Vec{x}_{k})\big)\big)}    
\end{equation}
where $f: \mathbb{R}^{2F+z}\rightarrow \mathbb{R}$.
We implement $f$ as a linear neural network.
The output of $f$ is subject to a non-linear activation function and subsequently normalized using softmax.
This transformation of $A$ allows for the learning of weights in the range $(0,1)$.
Since our method identifies and localizes anomalies, an edge weight $e_{ij}$ can be intuitively interpreted as the importance of a node $v_{j}$ to identify and localize an anomaly in node $v_{i}$ due to the fact that the node vector $\Vec{x}_{j}$ is scaled by the edge weight during convolution.
Thus, the actual convolution operation is defined as 
\begin{equation}
\Vec{x}'_i = \sum\limits_{j\in \altmathcal{N}(i)} \Tilde{A}_{ij}\cdot \Vec{x}_j.
\end{equation}
This spatial convolution allows for information transfer between neighboring nodes. 
After a subsequent ELU activation and residual connection, a final instance normalization is applied.

\subsection{Identification and Localization of Anomalies}
\label{sec:approach_classification}
The calculated node feature vectors of the graph are used for the final anomaly identification and localization block.
This is visualized in the last block of \autoref{fig:arvalus_step2}.
During training, another dropout layer is used to mitigate overfitting.
After that, a linear transformation maps from the node feature dimension to the output dimension of the model, where the output layer size equals the number of anomaly classes and the normal state class.
This effectively leads to a scalar value for each class. 
A subsequent softmax activation calculates a probability distribution over all classes, indicating the classification probability for each anomaly.
By investigating all nodes at the same time, we effectively combine anomaly identification and localization.

\section{Evaluation}
\label{sec:evaluation}

During evaluation we want to investigate the effect of modeling neighborhood inter-dependencies of cloud application components with graph convolution and its ability to identify and localize anomalies.
To evaluate our approach we simulate two cloud applications hosted within virtual machines of a cloud system and synthetically inject anomalies.
The initial graph represents the structure of two real cloud-hosted applications, a virtual IMS\footnote{\href{https://clearwater.readthedocs.io/en/stable/}{https://clearwater.readthedocs.io/en/stable/}} and a RTMP content streaming service (CS)\footnote{\href{https://github.com/arut/nginx-rtmp-module}{https://github.com/arut/nginx-rtmp-module}}, as well as their deployment on cloud virtual machines.
The KPIs of each system component are realized as samples drawn from differently parametrized distributions.
In our synthetic setup, this leads to 51 simulated system components for which we synthesize 10 KPIs each.
Subsequently, we simulate the injection of two anomaly types into the KPIs in order to obtain labeled data.
A summary of all system component states is summarized in \autoref{tbl:dataset_anomalies}. 
The absolute value of each drawn sample is used.
The anomaly is injected into exactly one randomly selected KPI of a target component $i$.
Furthermore, there is no temporal overlapping of injections, meaning that anomalies are expected to not run simultaneously.
Every anomaly is injected five times into each application service.
Taking into account~\autoref{tbl:dataset_anomalies} and selecting a random subset of $3\leq k \leq 7$ neighbors of a target system component $i$, for injection we randomly choose from three identified scenarios with probabilities 70\%, 20\% and 10\%:
\begin{itemize}
    \item \verb|Local|: While system component $i$ experiences the anomaly in its \emph{standard} form, all other nodes are normal.
    \item \verb|Neighborhood|: While the anomaly injected into system component $i$ reveals \emph{miscellaneous} characteristics, its respective $k$ neighbors are subject to an injection of the same anomaly in \emph{standard} form.
    \item \verb|Adversary|: While injecting an anomaly into system component $i$, its respective $k$ neighbors are subject to an injection of the corresponding \emph{adversary} anomaly. Both anomalies are in their respective \emph{standard} form.
\end{itemize}
This information is used used for labelling the components.
During identification and localization, the respective models will be trained to predict the node labels.


\begin{table}
\centering
\caption{List of artificial states (normal, anomalies), their parametrization and adversaries.}
\small
    \begin{tabular}[t]{c||c|c|c}
        \toprule
        \textbf{State} & \textit{Standard} & \textit{Miscellaneous} & \textit{Adversary}\\
        \midrule
        Anomaly 1 & $\mu=0, \sigma=0.1$ & $\mu=0, \sigma=0.17$ & Anomaly 2 \\
        Normal & $\mu=0, \sigma=0.2$ & - & - \\
        Anomaly 2 & $\mu=0, \sigma=0.3$ & $\mu=0, \sigma=0.23$ & Anomaly 1 \\
        \bottomrule
    \end{tabular}
\label{tbl:dataset_anomalies}
\end{table}

\subsection{Training Setup}
\label{sec:evaluation_setup}
In the following, we will describe the different aspects of our training setup.
First, we assign IDs to each injected anomaly.
We use a sliding window of size 20 and stride 20 on the simulated multivariate time series, slicing the dataset into subseries.
The subseries of all components are grouped to have identical logical start and end times.
We assign respective anomaly labels only if all samples in a series are drawn from the anomaly distribution.
Otherwise they are labeled as normal.
We employ a Leave-One-Group-Out (LOGO) cross-validator based on the assigned injection IDs.
Since every anomaly was injected five times into each application service, this results in five LOGO groups. 
For each KPI, its extracted subseries are preprocessed by rescaling to the range $(0, 1)$ with min-max normalization.
The min and max boundaries are determined per service group within the training dataset.

We parametrize our model as follows. 
For the node feature extraction, we employ three convolution filters for each service group with filter sizes $\{3,5,7\}$. 
By setting padding accordingly, the filters are configured to produce an output of the same length as the input subseries, i.e. 20 samples.
The same applies to the subsequent max-pooling operation, where a window of size 3 is used. 
With regards to biases, all linear layers in our architecture except the last one waive a learnable bias.
The same applies to the convolution layers.
The graph convolution layer of D-Arvalus introduces a limited number of additional parameters, dependent on the number of unique edge attribute vectors and the desired hidden model dimension.
For the model training, the improved softmax cross entropy loss function~\cite{Lin2017} with $\gamma=2$ is used to down-weight well-classified samples, such that more emphasis is put on the correct classification of hard samples. 
Further parameters are summarized in~\autoref{tbl:evaluation_setup}.

\begin{table}[ht]
\centering
\caption{Overview of model }
\small
    \begin{tabular}[t]{c|l}
        \toprule
        \textbf{Aspect}&\textbf{Configuration}\\
        \midrule
        Optimizer & Adam, $\text{lr}=10^{-2}$, $\beta_1=0.9,\ \beta_2=0.999$\\
        Training & Epochs = 100, batch size = 64\\
        \midrule
        \multirow{4}{*}{Model} 
        & In-Dim. = 20, Hidden-Dim. = 32, Out-Dim. = 3\\
        & Weight initialization: Xavier\\
        & \#Parameters: Arvalus (46.539), D-Arvalus (46.651)\\
        \midrule
        \multirow{3}{*}{Other} & weight decay = $10^{-5}$; Dropout (probability=50\%)\\
        & Instance normalization (without $\gamma$ and $\beta$)\\
        \bottomrule
    \end{tabular}
\label{tbl:evaluation_setup}
\end{table}
\subsection{Anomaly Identification}
\label{sec:evaluation_identification}
\begin{figure}
    \centering
    \includegraphics[width=\columnwidth]{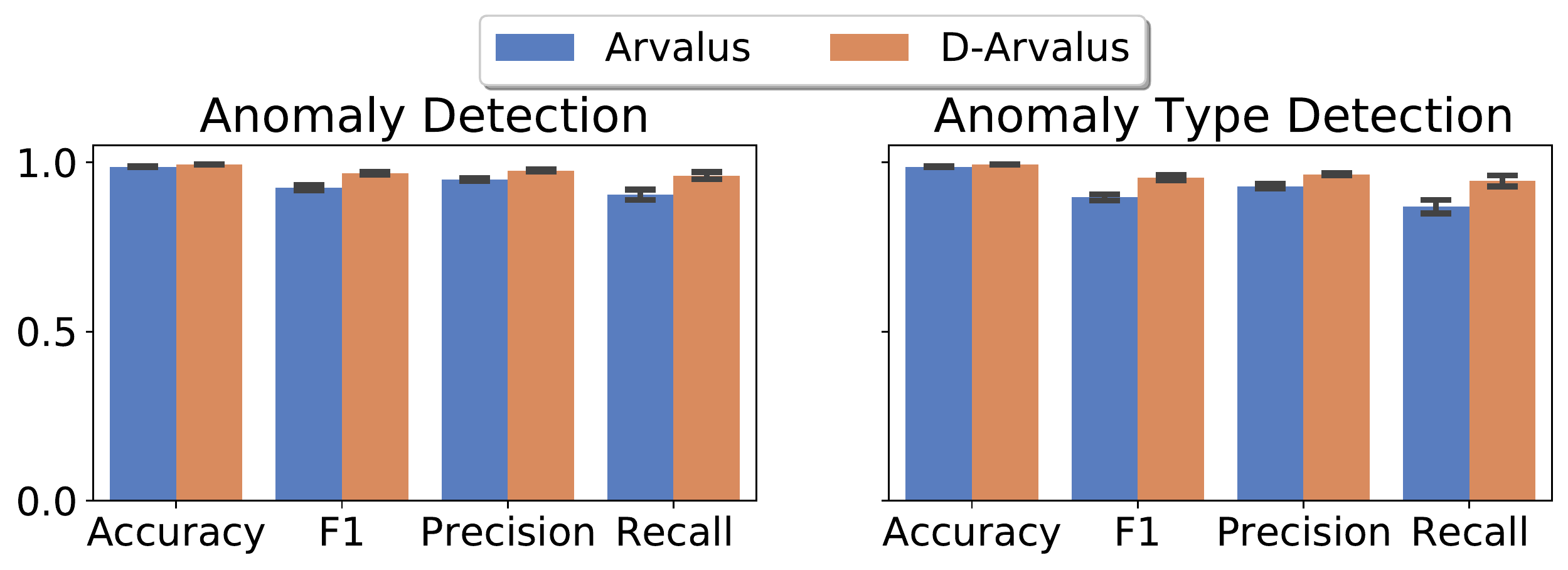}
    \caption{Identification of Anomalies via classification based on graph node features with (D-Arvalus) and without (Arvalus) graph convolution.}
    \label{fig:evaluation_detection_classification}
\end{figure}

\begin{figure}
    \centering
    \includegraphics[width=\columnwidth]{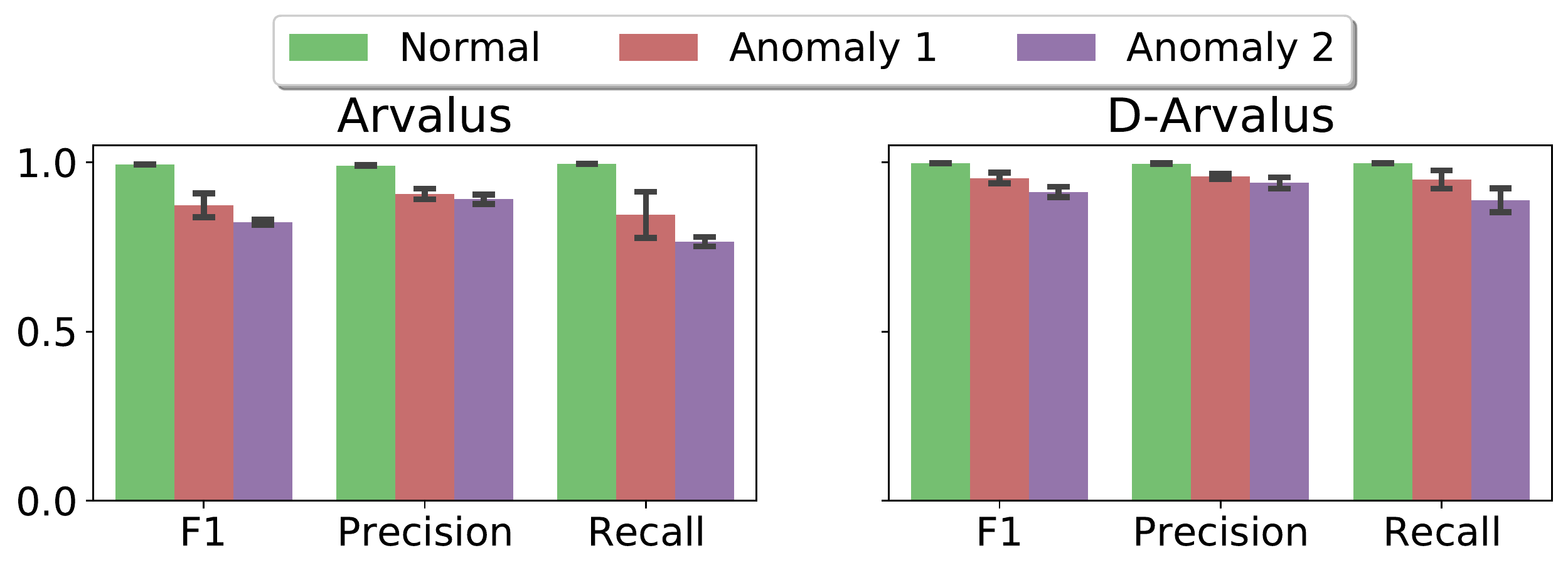}
    \caption{Insights into Detection of Anomaly Types: D-Arvalus allows for improved detection of anomaly classes}
    \label{fig:evaluation_classification_classes}
\end{figure}

To identify anomalies, classification on graph node features with (D-Arvalus) and without (Arvalus) graph convolution is applied.
We investigate accuracy, precision, recall and F1 scores.
For both, Arvalus and D-Arvalus, the validation result with the highest macro F1 score was selected for each split.
The results are depicted in \autoref{fig:evaluation_detection_classification} and \autoref{fig:evaluation_classification_classes}, where we present the average scores over all splits with corresponding standard deviation. 
\autoref{fig:evaluation_detection_classification} shows in general small standard deviation across splits. 
While the differences in accuracy appear insignificantly, the remaining evaluation scores are in general higher for D-Arvalus.
For anomaly detection, which is a binary classification task, it achieves an average macro F1 score of 0.967 compared to 0.925 of Arvalus. 
For anomaly identification, which is a multiclass classification task, a bigger gap between the F1 scores can be observed, i.e. 0.954 compared to 0.896. 
Due to the imbalance of classes in our scenario, we further unravel the results per class for more insights. 
\autoref{fig:evaluation_classification_classes} demonstrates that on average, D-Arvalus outperforms Arvalus in all evaluation metrics for all classes. 
Furthermore, the results are less volatile, as indicated by the small standard deviation across splits. 
While high scores are to be expected for the class representing the normal state, a performance increase for the count-wise underrepresented anomaly classes can be observed as well. 
From that, we conclude that learning the importance of dependencies between interconnected system components via our GC increases the evaluation scores in this evaluation setup.

\subsection{Anomaly Localization}
\label{sec:evaluation_localization}

\begin{figure}
    \centering
    \includegraphics[width=\columnwidth]{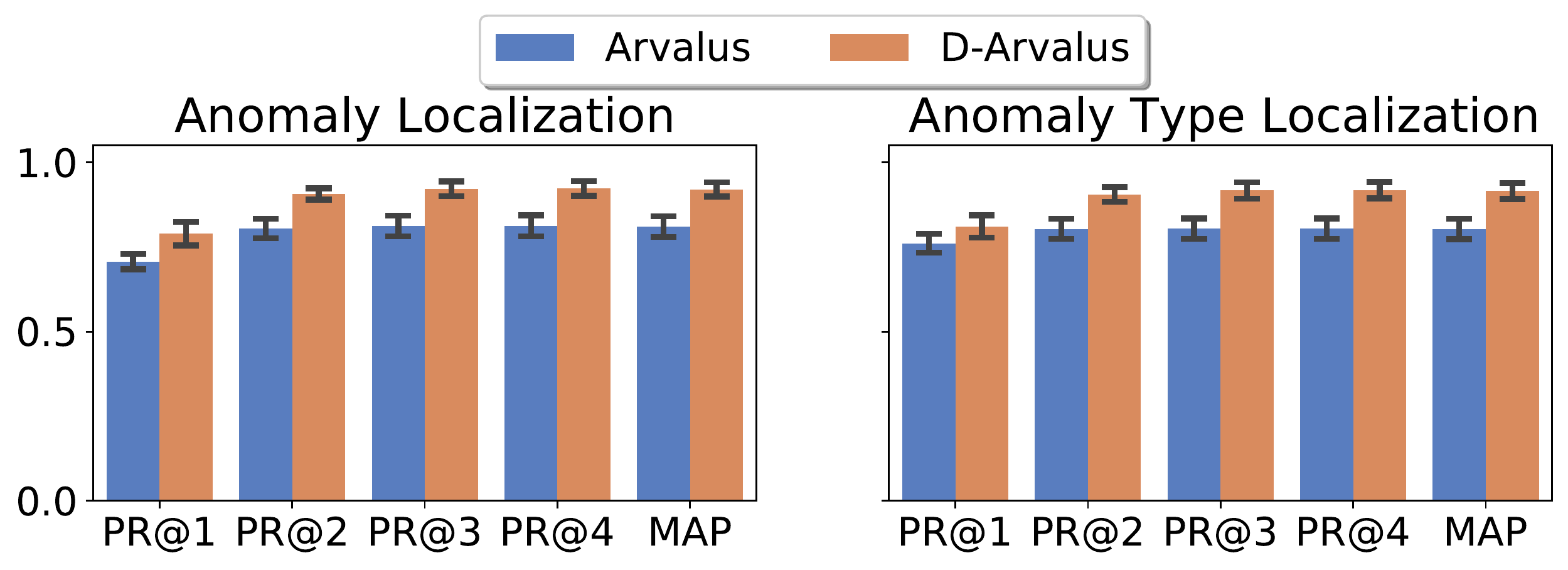}
    \caption{Localization of Anomalies: Higher localization precision due to D-Arvalus, as indicated by $MAP$ metric}
    \label{fig:evaluation_localization}
\end{figure}

To quantify the model performance on the task of anomaly localization based on a set of anomalies $\left | A \right |$ we use an adapted versions of the $PR@k$ and $MAP$ metrics~\cite{Wu2020microRCA}.
$PR@k$ defines the probability that the real anomaly location is among $k$ locations identified as potential locations.
A  high $PR@k$ score represents the algorithm's ability to correctly localize the anomaly in the system.
Smaller values of $k$ require the algorithm to be more precise to achieve high scores.
In practise, $k$ can be mapped to the worst case number of manual checks a human expert needs to perform in order to verify the localization of the anomaly.
Given $L$ as a set of predicted locations, $PR@k$ is defined as
\begin{equation}
    PR@k=\frac{1}{\left | A \right |} \sum_{a \in A} 
    \begin{cases}
        1, & \text{if } \left |L \right | \leq k\\
	    0, & \text{otherwise}
	\end{cases}.
\end{equation}
Consequently, for $N$ system components, $MAP$ is defined as  
\begin{equation}
    MAP = \frac{1}{N} \sum\limits_{1\leq j \leq N} PR@j
\end{equation}
Again, we compare the scores of Arvalus and D-Arvalus.
The anomaly localization results are presented in \autoref{fig:evaluation_localization}.
Anomaly localization measures the ability to localize the anomaly independent of of its concrete class label.
Anomaly type localization refers to the ability to localize the anomaly and predict the correct anomaly type.
Obviously, it can be observed that D-Arvalus outperforms Arvalus.
Its scores are higher even for small values of $k$. 
This performance gap is constant, as indicated by the $MAP$ results. 
On average, the $MAP$ score of D-Arvalus is $11 \%$ higher compared to Arvalus, and this can be shown both for anomaly localization and anomaly type localization. 
We further evaluate our previously defined scenarios and examine the accuracy of correctly localizing respective target nodes. 
Arvalus is on average able to correctly localize 97.6\% of \verb|Local|, 97.7\% of \verb|Adversary|, and 56.6\% of \verb|Neighborhood| target nodes. 
In contrast, D-Arvalus achieves results of 99.9\%, 99.0\% and 97.6\%. 
This demonstrates increased anomaly localization capabilities when incorporating neighborhood information.

In general, our approach leads to promising results in our simulated scenario, which motivates for future experiments using realistic distributed system testbeds~\cite{Nedelkoski20}.

\section{Conclusion}
\label{sec:conclusion}
In this paper we presented \emph{Arvalus}, a data-driven and graph-based approach for identifying and localizing anomalies in distributed cloud applications.
It allows to aggregate KPIs of interdependent application services by applying graph convolution (D-Arvalus) eventually resulting in node graph features that improve the identification and localization results.
Therefore, we developed a graph convolution method to model cloud application inter-dependencies.
Our evaluation experiments on a set of simulated cloud application KPIs shows that applying graph convolution consistently leads to better identification (Macro F1 score of 0.95 compared to 0.9) and localization (D-Arvalus has $11\%$ higher $MAP$ than Arvalus) results.
Thus, the graph convolution of D-Arvalus has the potential to improve many existing solutions.
In the future, we are planning to evaluate our method with data from real-world cloud deployments.

\section*{Data Availability}
We are pleased to make further technical details, experiment and evaluation scripts, as well as the used data publicly available\footnote{\href{https://doi.org/10.5281/zenodo.4589255}{https://doi.org/10.5281/zenodo.4589255}} to the community.

\section*{Acknowledgments}
This work has been supported through grants by the German Ministry for Education and Research (BMBF) as BIFOLD (funding mark 01IS18025A).

\bibliographystyle{IEEEtran}
\bibliography{bib}

\begin{thebibliography}{10}
\providecommand{\url}[1]{#1}
\csname url@samestyle\endcsname
\providecommand{\newblock}{\relax}
\providecommand{\bibinfo}[2]{#2}
\providecommand{\BIBentrySTDinterwordspacing}{\spaceskip=0pt\relax}
\providecommand{\BIBentryALTinterwordstretchfactor}{4}
\providecommand{\BIBentryALTinterwordspacing}{\spaceskip=\fontdimen2\font plus
\BIBentryALTinterwordstretchfactor\fontdimen3\font minus
  \fontdimen4\font\relax}
\providecommand{\BIBforeignlanguage}[2]{{%
\expandafter\ifx\csname l@#1\endcsname\relax
\typeout{** WARNING: IEEEtran.bst: No hyphenation pattern has been}%
\typeout{** loaded for the language `#1'. Using the pattern for}%
\typeout{** the default language instead.}%
\else
\language=\csname l@#1\endcsname
\fi
#2}}
\providecommand{\BIBdecl}{\relax}
\BIBdecl

\bibitem{wang2018cloudranger}
P.~Wang, J.~Xu, M.~Ma, W.~Lin, D.~Pan, Y.~Wang, and P.~Chen, ``Cloudranger:
  Root cause identification for cloud native systems,'' in \emph{CCGRID}.\hskip
  1em plus 0.5em minus 0.4em\relax IEEE, 2018.

\bibitem{dragoni2017microservices}
N.~Dragoni, S.~Giallorenzo, A.~L. Lafuente, M.~Mazzara, F.~Montesi,
  R.~Mustafin, and L.~Safina, ``Microservices: yesterday, today, and
  tomorrow,'' in \emph{Present and ulterior software engineering}.\hskip 1em
  plus 0.5em minus 0.4em\relax Springer, 2017.

\bibitem{Gulenko2016}
A.~Gulenko, M.~Wallschl{\"a}ger, F.~Schmidt, O.~Kao, and F.~Liu, ``A system
  architecture for real-time anomaly detection in large-scale nfv systems,''
  \emph{Procedia Computer Science}, 2016.

\bibitem{mariani2018localizing}
L.~Mariani, C.~Monni, M.~Pezz{\'e}, O.~Riganelli, and R.~Xin, ``Localizing
  faults in cloud systems,'' in \emph{ICST}.\hskip 1em plus 0.5em minus
  0.4em\relax IEEE, 2018.

\bibitem{Nedelkoski2019}
S.~Nedelkoski, J.~Cardoso, and O.~Kao, ``Anomaly detection from system tracing
  data using multimodal deep learning,'' in \emph{CLOUD}.\hskip 1em plus 0.5em
  minus 0.4em\relax IEEE, 2019.

\bibitem{sauvanaud2016anomaly}
C.~Sauvanaud, K.~Lazri, M.~Ka{\^a}niche, and K.~Kanoun, ``Anomaly detection and
  root cause localization in virtual network functions,'' in
  \emph{ISSRE}.\hskip 1em plus 0.5em minus 0.4em\relax IEEE, 2016.

\bibitem{li2019generic}
Z.~Li, C.~Luo, Y.~Zhao, Y.~Sun, K.~Sui, X.~Wang, D.~Liu, X.~Jin, Q.~Wang, and
  D.~Pei, ``Generic and robust localization of multi-dimensional root causes,''
  in \emph{ISSRE}.\hskip 1em plus 0.5em minus 0.4em\relax IEEE, 2019.

\bibitem{Wu2020microRCA}
L.~Wu, J.~Tordsson, E.~Elmroth, and O.~Kao, ``Microrca: Root cause localization
  of performance issues in microservices,'' in \emph{NOMS}.\hskip 1em plus
  0.5em minus 0.4em\relax {IEEE}, 2020.

\bibitem{Ahmed2016ASO}
M.~Ahmed, A.~Mahmood, and J.~Hu, ``A survey of network anomaly detection
  techniques,'' \emph{J. Netw. Comput. Appl.}, vol.~60, pp. 19--31, 2016.

\bibitem{zhou2019latent}
X.~Zhou, X.~Peng, T.~Xie, J.~Sun, C.~Ji, D.~Liu, Q.~Xiang, and C.~He, ``Latent
  error prediction and fault localization for microservice applications by
  learning from system trace logs,'' in \emph{ESEC/FSE}.\hskip 1em plus 0.5em
  minus 0.4em\relax {ACM}, 2019.

\bibitem{meng2020localizing}
Y.~Meng, S.~Zhang, Y.~Sun, R.~Zhang, Z.~Hu, Y.~Zhang, C.~Jia, Z.~Wang, and
  D.~Pei, ``Localizing failure root causes in a microservice through causality
  inference,'' in \emph{IWQoS}.\hskip 1em plus 0.5em minus 0.4em\relax IEEE,
  2020.

\bibitem{10.1145/2494232.2465753}
M.~Kim, R.~Sumbaly, and S.~Shah, ``Root cause detection in a service-oriented
  architecture,'' \emph{SIGMETRICS Perform. Eval. Rev.}, vol.~41, no.~1, p.
  93–104, Jun. 2013.

\bibitem{wang2019grano}
H.~Wang, P.~Nguyen, J.~Li, S.~Kopru, G.~Zhang, S.~Katariya, and
  S.~Ben-Romdhane, ``Grano: Interactive graph-based root cause analysis for
  cloud-native distributed data platform,'' \emph{Proceedings of the VLDB
  Endowment}, vol.~12, no.~12, pp. 1942--1945, 2019.

\bibitem{gulenko2018detecting}
A.~Gulenko, F.~Schmidt, A.~Acker, M.~Wallschl{\"a}ger, O.~Kao, and F.~Liu,
  ``Detecting anomalous behavior of black-box services modeled with
  distance-based online clustering,'' in \emph{CLOUD}.\hskip 1em plus 0.5em
  minus 0.4em\relax IEEE, 2018.

\bibitem{Zhang2019}
S.~Zhang, H.~Tong, J.~Xu, and R.~Maciejewski, ``Graph convolutional networks: a
  comprehensive review,'' \emph{Computational Social Networks}, vol.~6, no.~1,
  p.~11, 2019.

\bibitem{velivckovic2018graph}
P.~Velickovic, G.~Cucurull, A.~Casanova, A.~Romero, P.~Li{\`{o}}, and
  Y.~Bengio, ``Graph attention networks,'' in \emph{ICLR}.\hskip 1em plus 0.5em
  minus 0.4em\relax OpenReview.net, 2018.

\bibitem{wu:hal-02968710}
L.~Wu, J.~Tordsson, A.~Acker, and O.~Kao, ``{MicroRAS: Automatic Recovery in
  the Absence of Historical Failure Data for Microservice Systems},'' in
  \emph{UCC}.\hskip 1em plus 0.5em minus 0.4em\relax {IEEE}, 2020.

\bibitem{Lin2017}
T.~Lin, P.~Goyal, R.~B. Girshick, K.~He, and P.~Doll{\'{a}}r, ``Focal loss for
  dense object detection,'' \emph{{IEEE} Trans. Pattern Anal. Mach. Intell.},
  vol.~42, no.~2, 2020.

\bibitem{Nedelkoski20}
S.~Nedelkoski, J.~Bogatinovski, A.~K. Mandapati, S.~Becker, J.~Cardoso, and
  O.~Kao, ``Multi-source distributed system data for ai-powered analytics,'' in
  \emph{ESOCC}.\hskip 1em plus 0.5em minus 0.4em\relax Springer, 2020.

\end{thebibliography}

\end{document}